\documentclass[twocolumn,aps,preprintnumbers,prl,showpacs,amsfonts,amsmath,amssymb,superscriptaddress,floatfix]{revtex4}

\usepackage[dvips]{graphicx}
\allowdisplaybreaks[3]

\newcommand{\affA}{%
	Department of Applied Physics, School of Engineering, 
        The University of Tokyo,\\
	7-3-1 Hongo, Bunkyo-ku, Tokyo 113-8656, Japan}

\newcommand{\affB}{%
	CREST, Japan Science and Technology Agency, 
	1-9-9 Yaesu, Chuo-ku, Tokyo 103-0028, Japan}

\newcommand{\affC}{%
    Institut f\"{u}r Optik, Information und Photonik,
    Max-Planck Forschungsgruppe,\\
    Universit\"{a}t Erlangen-N\"{u}rnberg,
    G\"{u}nther-Scharowsky Str. 1, 91058 Erlangen, Germany}
    
\newcommand{\affD}{%
    Department of Physics, Technical University of Denmark, Building 309, 2800 Lyngby, Denmark}

\begin{document} 


\title{Demonstration of deterministic and high fidelity squeezing of quantum information}

\date{\today}

\author{Jun-ichi Yoshikawa}
\affiliation{\affA}
\affiliation{\affB}
\author{Toshiki Hayashi}
\author{Takayuki Akiyama}
\affiliation{\affA}
\author{Nobuyuki Takei}
\affiliation{\affA}
\affiliation{\affB}
\author{\\Alexander Huck}
\affiliation{\affA}
\affiliation{\affC}
\affiliation{\affD}
\author{Ulrik L.\ Andersen}
\affiliation{\affC}
\affiliation{\affD}
\author{Akira Furusawa}
\affiliation{\affA}
\affiliation{\affB}


\begin{abstract}
By employing at recent proposal (R. Filip, P. Marek and U.L. Andersen, Phys. Rev. A {\bf 71}, 042308 (2005)~\cite{Filip05.pra}), we experimentally demonstrate a universal, deterministic and  high-fidelity squeezing transformation of an optical field. It relies only on linear optics, homodyne detection, feedforward and an ancillary squeezed vacuum state, thus direct interaction between a strong pump and the quantum state is circumvented. We demonstrate three different squeezing levels for a coherent state input. This scheme is highly suitable for the fault-tolerant squeezing transformation in a continuous variable quantum computer.
\end{abstract}

\pacs{03.67.Lx, 03.67.Mn, 42.50.Dv}

\maketitle



The implementation of a direct nonlinear quantum operation is often hampered by decoherence due to inevitable practical imperfections in physical systems. Because of the necessity of invoking such unitary transformations in a fault-tolerant quantum information processor, the future of developing such units was not too bright. However, new optimism arose from the introduction of the so-called off-line schemes, where a nonlinear transformation is executed on a quantum state through simple linear interference with some off-line prepared ancillas followed by detection and feedforward~\cite{Gottesman99.nature,KLM,gottesman,Bartlett02.pra,Bartlett03.prl}. The significance of that approach is that the nonlinear transformation need not be performed directly onto the fragile quantum state, but is accomplished by tailoring the off-line resource states that can be prepared at anytime.

The first simple example of such an off-line scheme is teleportation which demonstrates the implementation of the most trivial unitary quantum operation~\cite{teleportation} - namely the identity operation: The off-line resource is a bipartite entangled state which is detected jointly with the fragile quantum information in a Bell measurement and the classical outcomes are fed forward to finalise the identity (or teleportation) operation. Remarkably it was found that by manipulating the off-line entangled state in the teleporter it is possible to implement any unitary transformation through teleportation. This was first realised for qubits~\cite{Gottesman99.nature} and subsequently used in the linear optical quantum computer~\cite{KLM}, and later extended to continuous variables (CVs) which benefit from the easy Bell measurement~\cite{Bartlett03.prl}. 

Such a teleportation-based off-line scheme can for example be used for the implementation of a unitary and nonlinear squeezing operation. It was however realised in ref.~\cite{Filip05.pra} that a much simpler off-line scheme relying only on a single vacuum squeezed ancilla suffices to implement the squeezing operation (see Fig. 1a). In essence, this simple setup allows for the experimentally feasible fault-tolerant squeezing transformation of {\it quantum information}, and it can be seen as the CV analog to the one-qubit teleportation approach in ref.~\cite{Zhou00.pra}. 
  
In this Letter we construct a squeezing transformation using the off-line approach proposed in Ref.~\cite{Filip05.pra}, and we demonstrate its function with coherent state inputs. 
Such a transformation is ideally described by a single mode Bogoliubov transformation, which maps the input Wigner function $W(x,p)$ onto $W^\prime(x,p)=W(xe^r,pe^{-r})$~\cite{Leonhardt} where $x$ and $p$ represent the amplitude and phase quadrature of the field and $r$ is the squeezing factor. Although this simple transformation is standard in any text book on quantum optics, its experimental realisation for arbitrary inputs (that is quantum information) has remained extremely challenging. Previously demonstrated squeezing transformations have either been suffering from large decoherence (as is the case for fiber or cavity implementations), thus corrupting the fragile quantum information of a quantum state, or been using an input dependent nondeterministic approach~\cite{lance06.pra}.
In contrast to previous implementations, the squeezing transformation demonstrated in this Letter is deterministic and it processes quantum information with very high fidelity. It is therefore the first demonstration of a near fault tolerant squeezing transformation that could be used in CV quantum computation~\cite{Bartlett02.pra,lloyd,CV_cluster}.

\begin{figure}[tb]
\centering
\includegraphics[clip,scale=0.5]{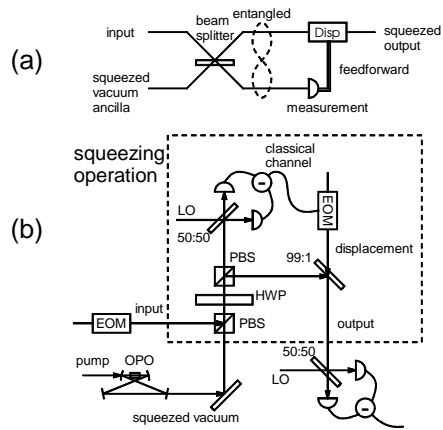}
\caption{
(a)~Schematic of high-fidelity squeezing. (b)~Experimental setup for high-fidelity squeezing. A variable beam splitter is realized by a half-wave plate (HWP) and two polarizing beam splitters (PBS). EOM: electro-optic modulator, LO: local oscillator, OPO: optical parametric oscillator.
}
\label{fig1}
\end{figure}

The scheme is illustrated in Fig.~\ref{fig1} and goes as follows. The input state under interrogation is combined with a squeezed vacuum at a beam splitter. A quadrature to be anti-squeezed is measured using homodyne detection, and after appropriate rescaling of the outcomes the remaining field is displaced accordingly. Mathematically, the transformation can easily be derived in the Heisenberg picture. First, we consider the input-output relations for the beam splitter:
\begin{align}
\hat{x}_\text{i}^\prime & =\sqrt{T}\hat{x}_\text{i}+\sqrt{1-T}\hat{x}_\text{a}, & 
\hat{p}_\text{i}^\prime & =\sqrt{T}\hat{p}_\text{i}+\sqrt{1-T}\hat{p}_\text{a},  \\
\hat{x}_\text{a}^\prime & =\sqrt{T}\hat{x}_\text{a}-\sqrt{1-T}\hat{x}_\text{i}, &
\hat{p}_\text{a}^\prime & =\sqrt{T}\hat{p}_\text{a}-\sqrt{1-T}\hat{p}_\text{i},
\end{align} 
where $\hat{x}$ and $\hat{p}$ represent the quadratures to be squeezed and anti-squeezed, the indices `i' and `a' refer to the input and ancillary mode, respectively, and $T$ is the transmittance of the beam splitter. The quadratures of the ancilla are written as $(\hat{x}_\text{a},\hat{p}_\text{a})=(\hat{x}_\text{a}^{(0)}e^{-r_\text{a}},\hat{p}_\text{a}^{(0)}e^{r_\text{a}})$ where $r_\text{a}$ is the squeezing parameter and $\hat{x}_\text{a}^{(0)}$ and $\hat{p}_\text{a}^{(0)}$ represent vacuum fluctuations. 
In the reflected part, the quadrature $\hat{p}_\text{a}^\prime$ is measured using homodyne detection. The measurement outcomes are subsequently rescaled by a factor denoted by $g$ and finally used to displace the remaining part of the system, which is equivalent to the transformation $\hat{x}_\text{i}^\prime\rightarrow \hat{x}_\text{i}^{\prime\prime}=\hat{x}_\text{i}^\prime$ and $\hat{p}_\text{i}^\prime\rightarrow \hat{p}_\text{i}^{\prime\prime}=\hat{p}_\text{i}^\prime+g\hat{p}_\text{a}^\prime$. By choosing $g=-\sqrt{(1-T)/T}$, we arrive at the following input-output relations
\begin{align}
\hat{x}_\text{i}^{\prime\prime} & =\sqrt{T}\hat{x}_\text{i}+\sqrt{1-T}\hat{x}_\text{a}^{(0)}e^{-r_\text{a}}, \label{sqz}\\
\hat{p}_\text{i}^{\prime\prime} & =\frac{1}{\sqrt{T}}\hat{p}_\text{i}.
\end{align}
In the limit of the infinitely squeezed ancilla, corresponding to $r_\text{a}\rightarrow \infty$, the transformation coincides with perfect unitary squeezing operation, with the actual squeezing parameter $r=-\ln\sqrt{T}$ which is directly controlled by the transmittance of the beam splitter. Furthermore, the quadrature being squeezed can also be easily controlled through adjustment of the relative phase between the signal and the squeezed ancilla and correspondingly the measured quadrature in the feedforward loop~\cite{Filip05.pra}. Therefore full control of the squeezing process is accessed through simple operations on linear passive devices. Let us note that by changing some of the settings of the setup (such as the local oscillator phase, the feedforward gain and the ancilla state) the setup can function as a non-unitary noiseless amplifier~\cite{lam}, a non-unitary  quantum nondemolition measurement device~\cite{buchler} or as a squeezed state purifier~\cite{glockl}.

In a realistic situation, the ancilla state is not infinitely squeezed and some extra quantum noise will inevitably be added to the squeezed quadrature as indicated by the second term in eq.~\eqref{sqz}. Note that the noise suppression performance never goes further than that of the ancilla. In contrast, the imperfections of the ancilla state do not degrade the quality of the transformation of the anti-squeezed quadrature as well as the mean values: The excess noise of the ancilla is not coupled into the mode nor does it disturb the mean value transformation.

The operation described above is universal and thus squeezes all input states. In the following experimental investigation, however, we consider the squeezing of particular states, namely coherent states. To ensure that the coherent states are truly pure, we define them to be a sideband at a radiofrequency relative to the carrier of a laser beam. This beam as well as other auxiliary beams are delivered by a Ti:Sapphire laser operating at 860~nm. The experiment is divided in three parts; preparation, processing and verification which will now be discussed.

{\it Preparation:} In the preparation stage, we generate the input coherent state and the squeezed ancilla state. The coherent state is prepared by traversing a part of the laser beam through an electro-optic modulator operating at 1~MHz and set to modulate the amplitude and phase simultaneously. As a result, a true coherent state is generated at a 1~MHz sideband and we assume the bandwidth to be 30~kHz. The power of the optical carrier is about 3~$\mu$W whereas the power of the sideband is about 15~dB above the corresponding shot noise level. 
The ancillary squeezed state is produced in an optical parametric oscillator (OPO). It is a 500~mm long bow-tie shaped cavity consisting of two plane mirrors and two mirrors with a 50~mm radius of curvature. The nonlinear crystal is a 10~mm periodically-poled KTiOPO{\scriptsize 4} (PPKTP) crystal (see~\cite{Suzuki06.apl} for details). We pump the OPO with light at 430~nm stemming from a second harmonic generator with the same configuration as the OPO cavity but with a KNbO{\scriptsize 3} crystal and pumped with the light from the Ti:Sapphire laser. To monitor and lock the squeezing phase we inject a weak coherent beam to the OPO. The output from the OPO and the coherent state are then directed to the processing part. They have 97~kHz and 143~kHz modulation sidebands for phase locking.

{\it Processing:} At this stage the actual squeezing transformation is implemented. First the two states from the preparation stage merge at a variable beam splitter composed of a half wave plate (HWP) sandwiched between two polarizing beam splitters (PBS). The beam splitting ratio is thus easily controlled via a wave plate rotation. One output of the beam splitter is directed to a homodyne detector which measures the $p$ quadrature. The visibility between the output and a local oscillator is 96\% and the quantum efficiency of the detectors is more than 99\%. The measurement outcomes are amplified electrically in a low-noise amplifier and subsequently used to drive a phase modulator which displaces an auxiliary beam. Finally, the displacement of the signal is achieved by combining it with the displaced auxiliary field using a highly asymmetric beam splitter (99/1).

{\it Verification:} In the final stage of the experiment, the protocol is verified by measuring the input states as well as the squeezed output states. The states are fully characterized by balanced homodyne detection. The visibility between the squeezed output beam and a local oscillator is 96\% and the total propagation efficiency is 96\%.
The electronic noise is always 19~dB smaller than the optical noise. After detection the photocurrents are used to reconstruct the quantum states: The 1~MHz component of the measured output signal is extracted by means of a lock-in detection scheme. The signal is mixed with a 1~MHz sine-wave signal from a function generator, low pass filtered (30~kHz) and finally digitized and fed into a computer with the sampling rate of 300~kHz.

\begin{figure}[tb]
\centering
\begin{tabular}{cc}
\includegraphics[clip,scale=0.36]{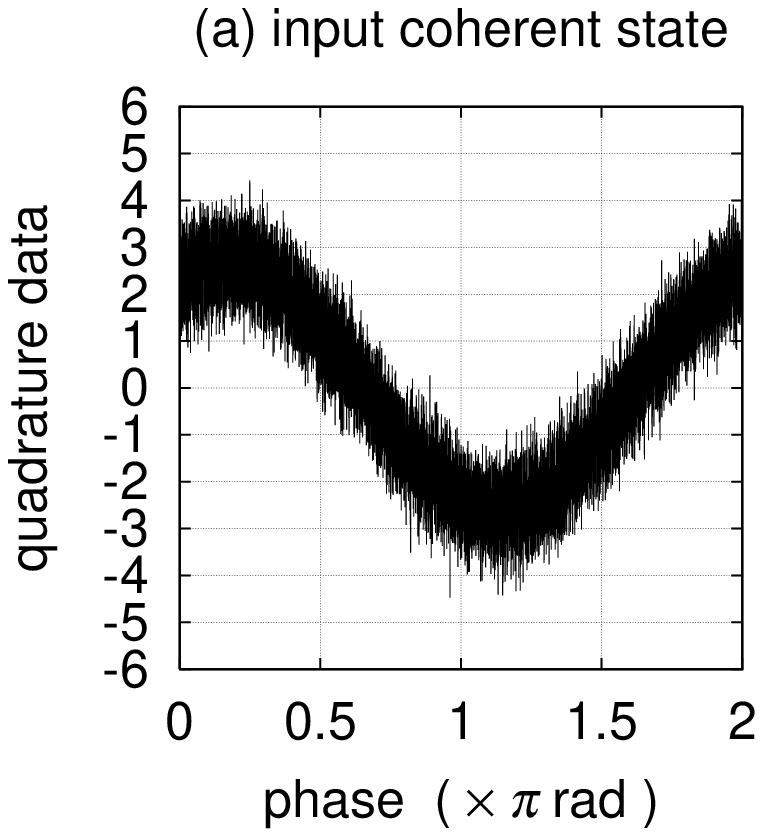} &
\includegraphics[clip,scale=0.36]{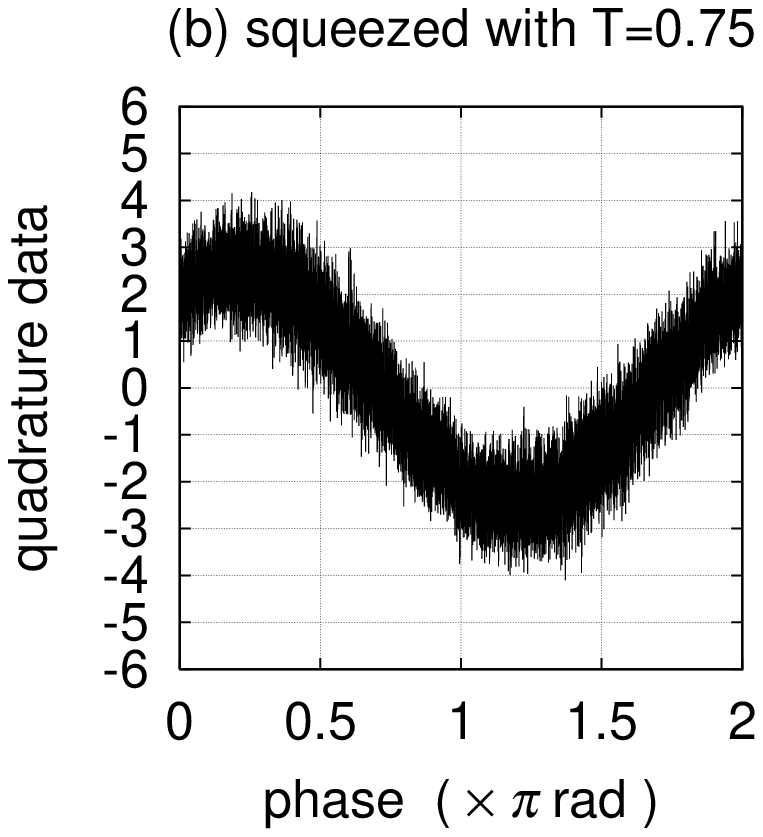} \\
\includegraphics[clip,scale=0.36]{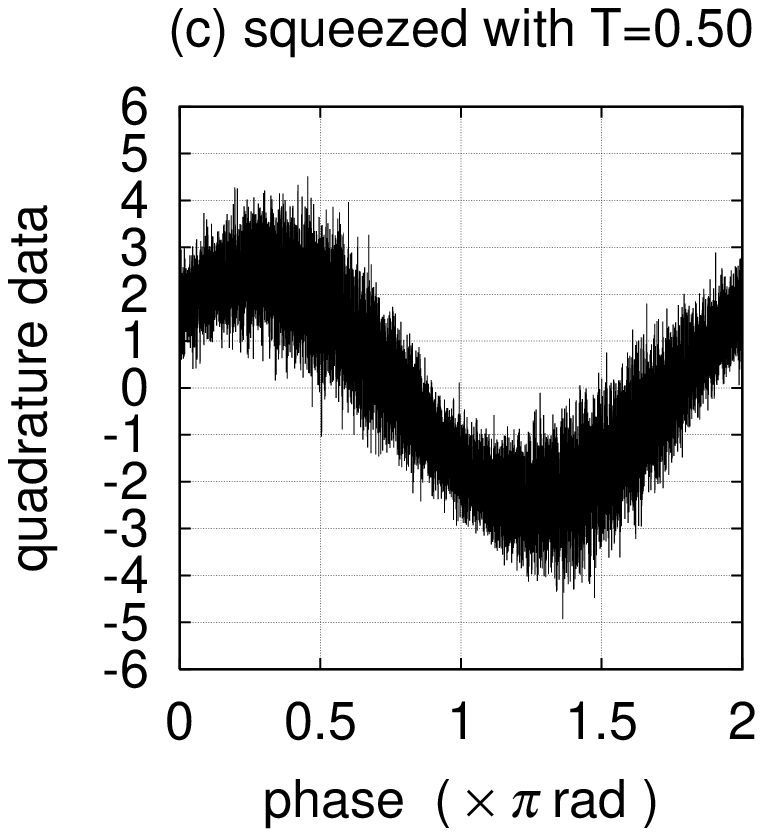} &
\includegraphics[clip,scale=0.36]{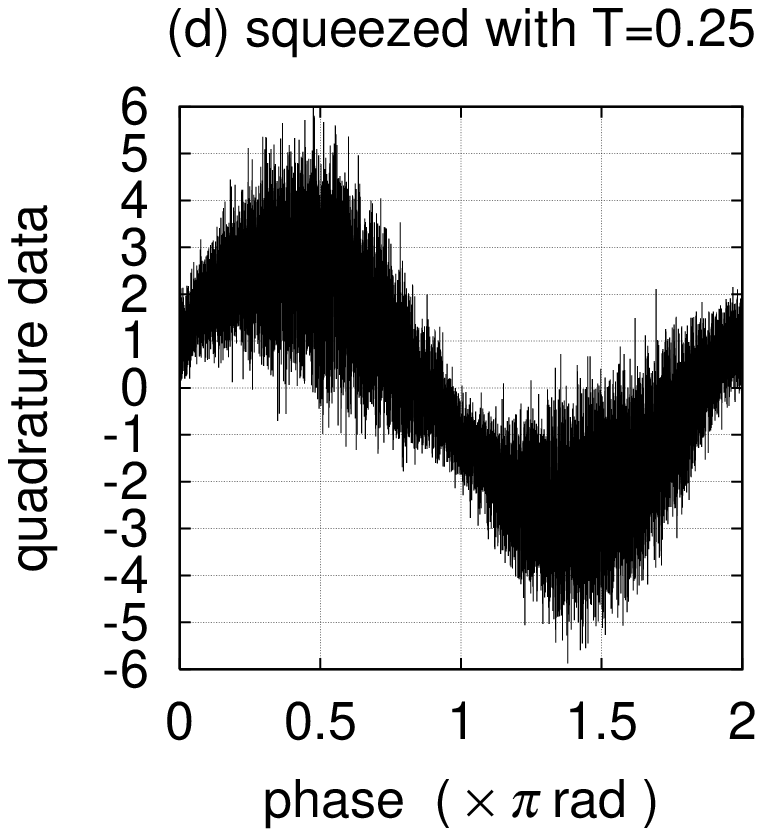} \\
\includegraphics[clip,scale=0.36]{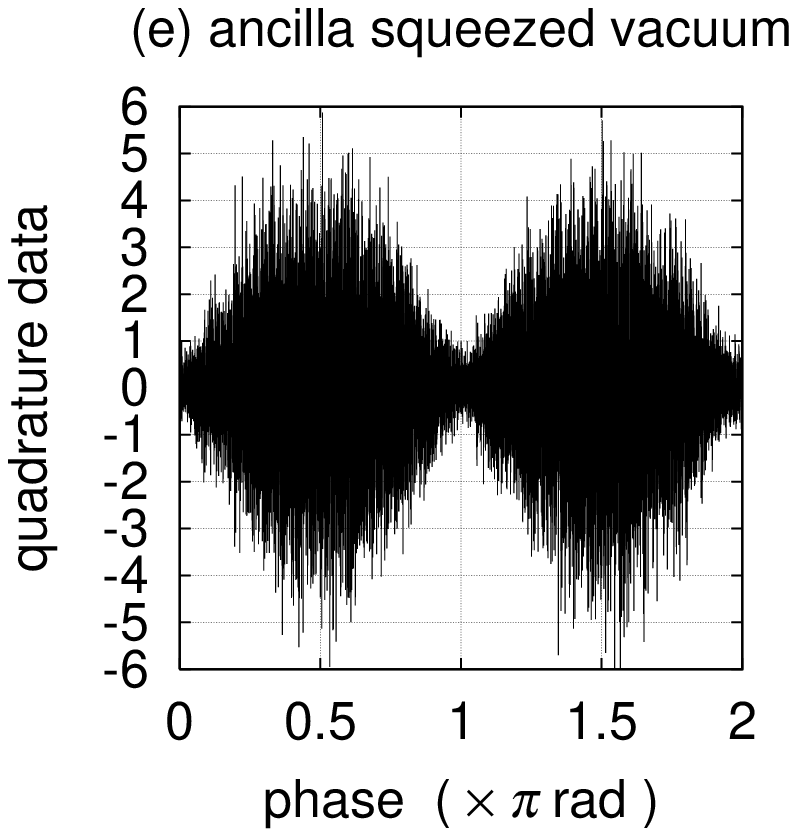} &
\includegraphics[clip,scale=0.54]{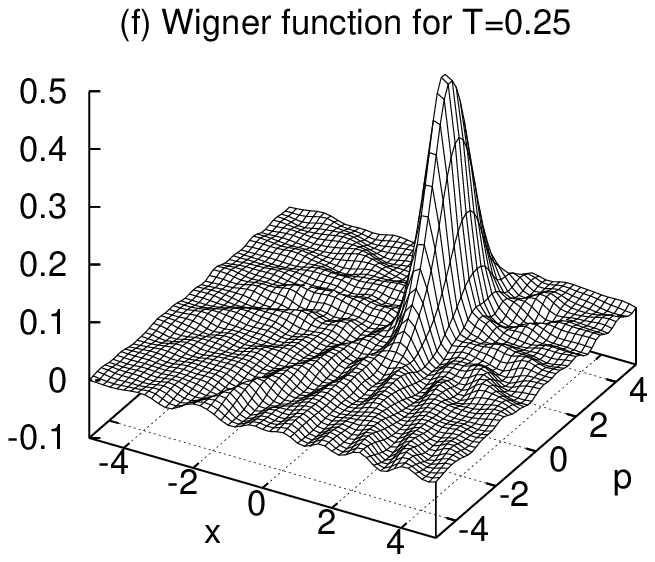}
\end{tabular}
\caption
{Results of the homodyne measurements. Fig. (a)-(e) are the raw quadrature data as a function of the phase of the local oscillator and (f) is the reconstructed Wigner function (using inverse Radon transformation~\cite{Leonhardt}) for one realization of the experiment.
}\label{fig2}
\end{figure}

First we present in Fig.~\ref{fig2} the raw data of the time resolved measurements of the input states and the output states. The time series for the input coherent states (Fig.~\ref{fig2}a) and the vacuum squeezed states (Fig.~\ref{fig2}e) are measured by adjusting the beam splitter transmittance to unity and zero, respectively (and blocking the displacement beam). We activate the squeezing transformation and measure the time series for three different transmittances, namely 0.75, 0.50, and 0.25, the results of which are shown in Fig.~\ref{fig2}b, \ref{fig2}c, and \ref{fig2}d respectively. It is evident from the plots that the input coherent states become more and more deformed as the transmittance decreases (and thus the squeezing degree increases). In Fig.~\ref{fig2}f, we present the reconstructed Wigner function of the transformed states with $T=0.25$.

\begin{figure}[tb]
\centering
\includegraphics[scale=0.56,clip]{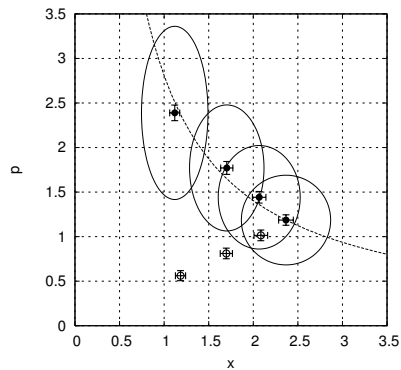}
\caption{Phase space representation of the experimental results.
The phase space is normalized so that the standard deviation of the vacuum fluctuation is $1/2$ ($\hbar=1/2$).
Note that the measured results are directly plotted without accounting for detection and propagation losses.
}\label{fig3}
\end{figure}

\begin{figure}[tb]
\centering
\includegraphics[scale=0.56,clip]{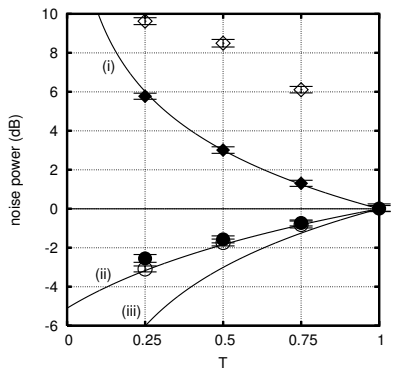}
\caption{
The noise powers of the squeezer outputs relative to the shot noise limit. We measure 0.7~dB, 1.6~dB and 2.5~dB for the squeezed quadrature and 1.3~dB, 3.0~dB and 5.8~dB for the anti-squeezed quadrature. The two quadrature variances of the input coherent states are equal to those of vacuum within $\pm$0.1~dB.
}\label{fig4}
\end{figure}

As indicated by the reconstructed Wigner function, the involved states are Gaussian. With this a priori information the states are completely characterized by its first two moments.
Due to the symmetry of the states (squeezed in $x$ and anti-squeezed in $p$) it suffices to evaluate the mean values and variances of $x$ and $p$. Results of such evaluations are shown in Fig.~\ref{fig3} and Fig.~\ref{fig4}.

In Fig.~\ref{fig3} the phase space diagrams of the input coherent states as well as the output states are shown by ellipses, which correspond to the cross sections of their respective Wigner functions. 
When a coherent state is unitarily squeezed the amplitude is transformed along a hyperbolic curve, as shown by the dotted line.
The four ellipses correspond to (from the right) the input coherent states, the squeezed outputs with $T=0.75$, $T=0.50$, $T=0.25$, respectively, and their centers, marked by dots, represent the measured averages.
The circles represent the data obtained without the feedforward.
The lengths of the major and minor axes of the ellipses are the measured standard deviations of $x$ and $p$. Obviously the mean values are transformed almost ideally.

In Fig.~\ref{fig4} the noise powers of the squeezed and anti-squeezed quadratures are plotted as a function of the transmittance. The three curves represent theoretical predictions for the noise power of the anti-squeezed quadrature (curve i), the squeezed quadrature with the ancilla 5.1~dB squeezed (curve ii) and infinitely squeezed (curve iii). Note again that the anti-squeezed noise does not depend on the ancilla.
Experimental data taken with and without the feedforward in place are also shown in Fig.~\ref{fig4}: The noise powers of $x(p)$ with feedforward are indicated by dots(filled diamonds), and without feedforward by circles(open diamonds).
We see that the anti-squeezed noise of the ancilla is cancelled and the transformation in $p$ becomes almost ideal after the feedforward. The noise powers of the squeezed qudrature, however, deviate from the ideal operation due to the finite squeezing in the ancilla states. Furthermore we observe a small degradation of the noise suppression due to some imperfections of the feedforward, such as phase fluctuation.

We now calculate the fidelities~\cite{Schumacher95.pra} of these transformations.
For the case of Gaussian states the fidelity between the ideal squeezed state, $|\psi_\text{id}\rangle$, and the actual obtained mixed state, $\hat{\rho}_\text{out}$, is given by (in the unit of $\hbar=1/2$),
\begin{align}
F= & \langle \psi_\text{id} | \hat{\rho}_\text{out} | \psi_\text{id} \rangle  \notag \\
 = & \frac{1}{2\sqrt{(V_\text{out}^x+V_\text{id}^x)(V_\text{out}^p+V_\text{id}^p)}}  \notag \\
   & \exp\left[-\frac{(\langle x_\text{out} \rangle -\langle x_\text{id} \rangle )^2}{2( V_\text{out}^x+V_\text{id}^x)}-\frac{(\langle p_\text{out} \rangle -\langle p_\text{id}\rangle)^2}{2(V_\text{out}^p+V_\text{id}^p)}\right],
\label{fidelity}
\end{align}
where the subscripts `id' and `out' denote the ideal squeezing and the experimental output, respectively, and $V$ denotes the variance.  
Actually, due to small propagation and detection losses in the experiment, the fidelity ultimately depends on the input state. We therefore quantify the individual single shot fidelities for the inputs considered in the experiment, though the average fidelity will be found by integrating the fidelity in eq.~\eqref{fidelity} over all possible input states. 
From the measured means and variances we compute the fidelities between the ideally squeezed states of the inferred inputs (accounting for losses) and the directly measured squeezed states, and we find 94\%$\pm$1\% for $T=0.75$ (1.2~dB squeezing), 89\%$\pm$1\% for $T=0.50$ (3.0~dB squeezing), and 78\%$\pm$2\% for $T=0.25$ (6.0~dB squeezing).
We note that the fidelity between the measured input states and the inferred ones is found to be 97\%$\pm$1\%.
For comparison, the theoretically calculated fidelities with vacuum ancilla states (which correspond to the classical limits) are 93\%, 82\%, 63\%, for the transformations corresponding to 1.2~dB, 3.0~dB, 6.0~dB squeezing, respectively.

In summary, we have succeeded in demonstrating deterministic and universal squeezing transformation using a feedforward technique. The squeezing operation associated with three different squeezing degrees corresponding to 1.2~dB, 3.0~dB, and 6.0~dB, were demonstrated and quantum noise suppressions of 0.7~dB, 1.6~dB and 2.5~dB below the shot noise were obtained, yielding the fidelities 94\%$\pm$1\%, 89\%$\pm$1\% and 78\%$\pm$2\%, respectively. Although the transformation only was tested for coherent states of one kind, it will work equally well for any other state due to its universality. 


Finally, we should note that the high-fidelity squeezer in this work completes the set of demonstrated Gaussian operations~\cite{lloyd,braunstein}. An arbitrary multi-mode Gaussian transformation can be physically generated by the use of phase space displacement and rotation, beam splitting interaction, phase insensitive amplification~\cite{josse06.prl} and universal squeezing transformation. With the work presented in this paper, we therefore pave the way for the experimental demonstration of new interesting CV Gaussian protocols such as the CV Controlled-NOT gate~\cite{Filip05.pra} and eventually quantum computation.

ULA and AH acknowledge financial support from the EU under project No. FP6-511004 COVAQIAL. This work was partly supported by the MEXT of Japan.

\end{document}